\documentclass[aps,prl,showpacs,twocolumn,groupedaddress,floatfix]{revtex4}
\usepackage{graphics}
\usepackage{graphicx}
\usepackage{epsfig}
\usepackage{pstricks}
\usepackage{color}

\usepackage{amsmath,amssymb} 
\usepackage{bm}  

\newcommand{\TBECZERO}{T^0_{\text{BEC}}}
\newcommand{\KTB}{\text{KT}}
\newcommand{\rhos}{\rho_{\text{s}}}
\newcommand{\rc}{r_{\text{c}}}
\newcommand{\nc}{n_{\text{c}}}
\newcommand{\nonc}{\text{nc}}

\newcommand{\mf}{\text{mf}}
\newcommand{\cl}{\text{cl}}
\newcommand{\half}{\tfrac{1}{2}}
\newcommand{\flabel}[1]{\label{f:#1}}
\newcommand{\elabel}[1]{\label{e:#1}}
\newcommand{\eq}[1]{Eq.~(\ref{e:#1})}

\newcommand{\fig}[1]{Fig.~\ref{f:#1}}
\newcommand{\figg}[1]{Figure~\ref{f:#1}}

\newcommand{\quot}[1]{``#1''}

\newcommand{\rvec}{{\bf r}}
\begin{document}

\title{Kosterlitz--Thouless transition of the quasi two-dimensional trapped Bose gas}

\author{Markus Holzmann}
\affiliation{ LPTMC,  Universit\'e Pierre et Marie Curie, 4 Place Jussieu,
75005 Paris, France; and LPMMC, CNRS-UJF,  BP 166, 38042 Grenoble, France}
\author{Werner Krauth}
\affiliation{CNRS-Laboratoire de Physique Statistique, Ecole Normale
Sup\'{e}rieure, 24 rue Lhomond, 75231 Paris Cedex 05, France}
\date{\today}

\begin{abstract}
We present Quantum Monte Carlo calculations with up to $N=576\,000$
interacting bosons in a quasi two-dimensional trap geometry closely
related to recent experiments with atomic gases. The density profile of
the gas and the non-classical moment of inertia yield intrinsic signatures
for the Kosterlitz--Thouless transition temperature $T_{\KTB}$.  From the
reduced one-body density matrix, we compute the condensate fraction, which
is quite large for small systems.  It decreases slowly with increasing
system sizes, vanishing in the thermodynamic limit.  We interpret our
data in the framework of the local-density approximation, and point out
the relevance of our results for the analysis of experiments.
\end{abstract}

\maketitle

Phase transitions in two-dimensional systems with a continuous
order parameter are of special interest because long-range order is
absent at finite temperatures, as a consequence of the Mermin--Wagner
theorem \cite{Hohenberg,Mermin}.  Instead, a Kosterlitz--Thouless phase
transition \cite{KT} can separate the high-temperature disordered phase
with exponential decay of the order parameter from a low-temperature
ordered phase with algebraically decaying order parameter, which was
proposed by Berezinskii \cite{Berezinskii}.

Recently, the Kosterlitz--Thouless phase transition was observed
in trapped quasi two-dimensional Bose gases of $^{87}$Rb atoms
\cite{Dalibard_2006,Dalibard_2007}, but the interpretation of the experimental
data was rendered difficult because of relatively strong interactions
in two dimensions, the trap confinement, and pronounced
finite-size corrections. Moreover, the ideal two-dimensional trapped
Bose gas is not a good vantage point to approach the weakly interacting
gas: the former Bose-condenses at  finite temperature (with a diverging
density in the center), whereas the latter becomes a superfluid, but
with vanishing condensate fraction.

Quantum Monte Carlo (QMC) calculations allow us to compute the thermodynamic
properties of interacting Bose systems \cite{ceperley} for a finite
number $N$ of particles, and for a wide range of microscopic interaction
parameters. The calculations are practically free of systematic errors. For
trapped atomic gases, very large particle numbers can be handled,  in
confining geometries relevant to current experiments \cite{qmc,mchf,g2}.
The ideal Bose gas enters these calculations in an exact way, and only
pair interactions give rise to the usual Metropolis rejection
process \cite{SMAC}.

In this paper, we present QMC calculations of three-dimensional
trapped bosons interacting with an $s$-wave pseudopotential in a
pancake-shaped harmonic trap with frequencies $\omega = \omega_x =
\omega_y \ll \omega_z$.  Our particle numbers range from $N=2000$
to   $N=576\,000$, the latter exceeding current experiments by more
than one order of magnitude.  The diagonal many-particle density
matrix directly yields the density profile and the non-classical
moment of inertia. Both allow us to locate the phase transition.
We also determine the condensate fraction explicitly from the largest
eigenvalue of the reduced off-diagonal one-particle density matrix. In a
trap, this calculation is more complicated than in a homogeneous system,
where the $k=0$ groundstate wavefunction is trivially known and where
the groundstate occupation governs the long-range behavior of the
off-diagonal one-particle density matrix.

The trapped ideal two-dimensional  Bose gas shows a Bose--Einstein
transition \cite{Kleppner} at a non-vanishing temperature $\TBECZERO =
\sqrt{6 N \omega^2 /\pi^2}$  (we choose units with $\hbar=m=\omega=1$,
where $m$ is the atomic mass). The central density diverges
logarithmically with $N$ (see \fig{profile_1}). This implies that
interaction effects play a much more pronounced role, even above the
transition temperature, than in three-dimensional traps.

As in the experiment \cite{Dalibard_2007}, we allow for a finite
extension $\omega_z$ of the trap in the $z$-direction, keeping the level
spacing on the order of the temperature: $ \omega_z = 0.55\,\TBECZERO$.
In the many-body density matrix, the $z$-dependence is dominated by a
(normalized) single-particle contribution, 
$\rho(z,z')$, which
separates out, and the effective two-dimensional interaction
strength is given by
\begin{equation}
g = \frac{4 \pi \hbar^2 a_0}{m} \int \text{d}z\, [\rho(z,z)]^2, 
\elabel{effective_two_d}
\end{equation}
where $a_0$  is the three-dimensional $s$-wave scattering length.
For particles distributed in $z$ according to the groundstate of the
harmonic oscillator, \eq{effective_two_d} reduces to $g =
\tilde{g} \equiv a_0 \sqrt{8 \pi \omega_z}$.  For our simulations, we
have used the experimental value $\tilde{g}= 0.13$ \cite{Dalibard_2007}, however,
the actual value of $g$ can be obtained directly from the computed density profile in the
$z$-direction.

We study the anisotropic trap at temperatures comparable to $\TBECZERO$
where it is indeed quasi two-dimensional because the extension in $z$ is
comparable to the deBroglie wavelength $\lambda=\sqrt{2\pi/mT}$.  
Formally, the three-dimensional Bose--Einstein transition
temperature $0.94\, N^{1/3}(\omega^2 \omega_z)^{1/3}$  is  of the same
order as $\TBECZERO$. However, the three-dimensional limit requires
that $\omega_z/\omega$ remains constant independent of the system size
in contrast to our quasi two-dimensional limit where $\omega_z/\omega \sim N^{1/2}$.
\begin{figure}
   \epsfig{figure=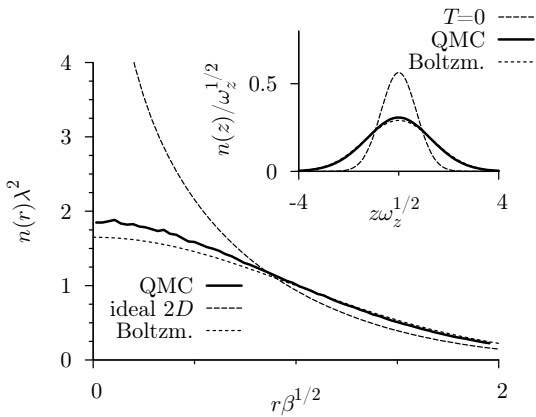,width=0.4\textwidth}
   \caption{Two-dimensional density profile $n(r) \lambda^2$ at
   $T=\TBECZERO$ for $N=576\,000$, compared to the saturation density of
   the ideal Bose gas and the density profile of an ideal gas of
   distinguishable particles.  The inset compares the density profile
   in $z$ to the groundstate distribution of the harmonic oscillator
   and to the ideal gas of distinguishable particles.}
\flabel{profile_1} 
\end{figure}

Recent numerical calculations \cite{Svistunov2D} have determined the
critical density $\nc$ at the Kosterlitz--Thouless transition
in the weakly interacting two-dimensional homogeneous Bose gas  of
density $n$,
\begin{equation}
   \nc \lambda^2 \simeq \ln \frac{380}{mg}.
   \elabel{svist}
\end{equation}
The interaction $g$ enters this expression only logarithmically, and the
differences between the actual $g$ and $\tilde{g}$, of the order of $40\%$
at $T_{\KTB}$, only results in a $6\%$ shift in the critical density.

In the trapped Bose gas, within the local-density approximation, the
transition takes place when the central density $n(0)$ equals the critical
density of the homogeneous gas, in our case $\nc(0) \lambda^2 \simeq 8$.
Mean-field theory \cite{TKTB} predicts that the Kosterlitz--Thouless
transition is somewhat below $\TBECZERO$:
\begin{equation}
  \frac{T_{\KTB}^{\mf}}{\TBECZERO}=\left(1+\frac{3 g}{\pi^3} 
  \left[ \nc(0)\lambda^2 \right]^2 \right)^{-\half}.
  \elabel{tckt}
\end{equation}
The mean-field value of $T_{\KTB}$, together with 
the numerical value of \eq{svist} for the critical density in
the center of the trap allows to determine the critical
temperature of the Kosterlitz--Thouless transition in the trap,
in our case $T_{\KTB}^{\mf}\simeq 0.75\,\TBECZERO$.

In \fig{profile_1}, we show the two-dimensional density profile $n(r)$
with $r=\sqrt{x^2+y^2}$
from our QMC calculations 
at $T=\TBECZERO$ for $N=576\,000$.
We also
illustrate the large deviations from the saturation density of the
two-dimensional ideal Bose gas \cite{SMAC}. Indeed, the density profile is
closer to that of ideal quantum Boltzmann particles described by the
density matrix of the harmonic oscillator.  The  density is everywhere
below the critical value, confirming that the interacting gas remains in
its high-temperature phase at lower temperatures than the ideal Bose gas.

\begin{figure}
\epsfig{figure=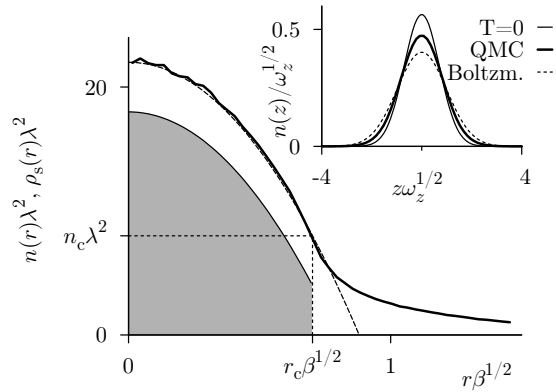,width=0.4\textwidth}
   \caption{Two-dimensional density profile $n(r) \lambda^2$ at
   temperature $T = 0.5\,\TBECZERO$ for $N=576\,000$ (thick line), compared to the
   Thomas--Fermi profile of \eq{Thomas_Fermi} (with $g = 0.109$, dashed line). The
   ansatz of \eq{ansatz} for the superfluid density $\rhos(r)$, with
   the universal jump at $r=\rc$, corresponds to the shaded region.
    The inset compares the
   density profile in the tightly confined $z$-direction to the
   groundstate distribution of the harmonic oscillator and the
   distribution of an ideal gas of distinguishable particles. }
   \flabel{profile_0.5}
\end{figure}

In \fig{profile_0.5}, we show the analogous density profile at temperature
$T/\TBECZERO = 0.5$, again for $N=576\,000$ particles. The central
density is now well in excess of the critical value of \eq{svist}.
We may define a \quot{critical radius} $r_{\text{c}}$, which  separates the
\quot{inner region} of the trap, with $r< \rc$ and $n(r) > \nc$,
from an \quot{outer region} with $r > \rc$ and with $n(r) < \nc$.  In the
local-density approximation, the inner region  is in the superfluid phase,
whereas the outer region is normal. At the critical radius, 
the density is at the Kosterlitz--Thouless phase-transition temperature.
In the inner region, $n(r)$
is very  well described by a Thomas--Fermi profile
\begin{equation}
n(r) = n(0) - \frac{\omega^2 r^2}{2 g}, \quad r < \rc,
\elabel{Thomas_Fermi}
\end{equation}
with the effective two-dimensional interaction parameter at this
temperature $g=0.107$, obtained, via \eq{effective_two_d}, directly
from the density profile in $z$ (see the inset of \fig{profile_0.5}).
The latter is wider than the groundstate distribution of the harmonic
oscillator, so that  $g$ is smaller than $\tilde{g}$.  The density profile
in $r$, whose width depends linearly on $g^{-1}$, is more sensitive
to the detailed value of the interaction than the transition temperature,
which decreases with the  logarithm of $g$.

In \fig{curvature}, we plot the central density $n(0) \lambda^2$ and also
the (central) curvature $\kappa=-(\lambda^2/\beta)\partial n(r) /\partial (r^2)|_{r=0}$.
The curvature of the Thomas--Fermi profile (in \eq{Thomas_Fermi}) is 
$\kappa = \omega^2 \pi/g$
with $g=\tilde{g}$ at very low temperature.  The curvature increases (the
profile becomes narrower) with $T$ because particles spread out farther
in the $z$-direction.  Above the critical temperature, however, the curvature
decreases (the profile becomes wider), as is natural for a thermal gas,
with  $\kappa \propto n(0)\lambda^2$.  The curvature plot provides an
intrinsic signature of the phase transition, at a temperature $T/\TBECZERO
\simeq 0.70$, which agrees nicely with the temperature at which the
central density passes the critical value \eq{svist}. 
We have also studied smaller
systems (with $N=2250, 9000, 36\,000$, and $144\,000$) at
unchanged values of $T/\TBECZERO$ and $\omega_z/\TBECZERO$,
but found only
very small variations in the density profiles.
Our value for the
critical temperature, $T_{\KTB}\simeq 0.70\,\TBECZERO$ is  close to the mean-field formula of \eq{tckt},
and somewhat higher than the experimental value \cite{Dalibard_2007}
$T_{\KTB}^{\text{exp}} \simeq  0.46(3)\,\TBECZERO$.  
\begin{figure}
   \epsfig{figure=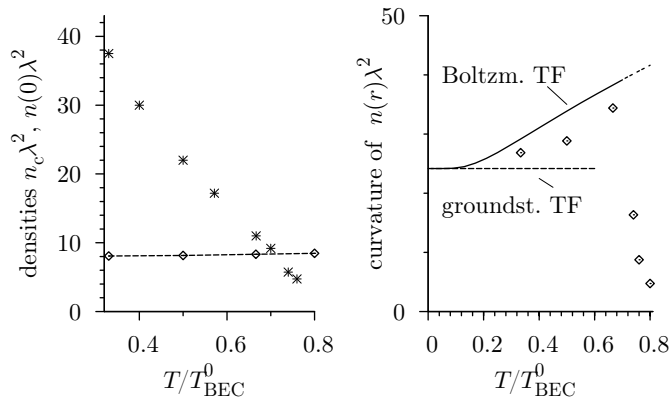,width=0.5\textwidth}
   \caption{{\emph{Left}: Densities $\nc \lambda^2$ and $n(0)\lambda^2$
   \emph{vs.} $T/\TBECZERO$ for the quasi two-dimensional gas with
   $N=576\,000$. The intersection of both curves leads to a 
   transition temperature $T_{\KTB} \simeq 0.70$.}
    \emph{Right}: central curvature $\kappa$ compared to the groundstate
    Thomas--Fermi curvature, \eq{Thomas_Fermi}, with $g=\tilde{g}$
   and with $g$
   corresponding to an ideal gas of distinguishable particles. The
   central curvature changes slope at $T_{\KTB}$.} 
   \flabel{curvature}
\end{figure}

The low-temperature phase below the Kosterlitz--Thouless transition is
a superfluid.  For a homogeneous system, the superfluid fraction can be
probed through the response to boundary conditions, and easily computed
within the  path-integral formalism, through the winding-number formula
\cite{ceperley}.  Likewise, a trapped superfluid does not respond to
an infinitely slow rotation of a trap leading to a non-classical
moment of inertia, $I_{\nonc }$, which is smaller than the classical value
$I_{\cl}= \int \text{d}\rvec \, r^2 n(r)$.  The non-classical moment
of inertia can again be computed from the diagonal elements of the
density matrix \cite{Sindzingre}.  In a homogeneous system, the ratio of
the
non-classical moment to the classical moment equals  the normal fraction.
In \fig{inertia}, we show that
a superfluid phase emerges below $T \simeq 0.70\,\TBECZERO$, and that
$I_{\nonc}/I_{cl}$ remains different from unity, independent
on system size.
\begin{figure}
   \epsfig{figure=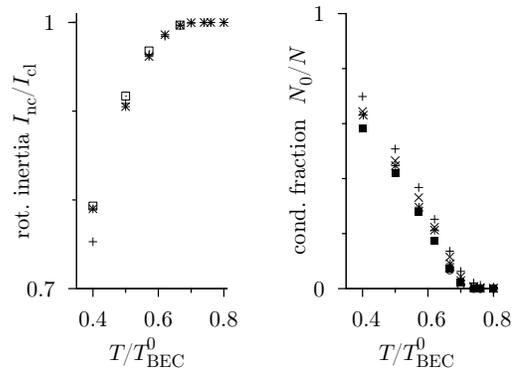,width=0.4\textwidth}
   \caption{\emph{Left:} Non-classical moment of inertia 
   $I_{\nonc}/I_{\cl}$ \emph{vs.} $T/ \TBECZERO$ for
   $N=9000$ (crosses) and $N=144\,000$ (stars) compared
   to the ansatz of \eq{ansatz} (squares). 
   \emph{Right:} Condensate fraction for particle numbers ranging from 
   $N=2250$ (crosses) to $N=144\,000$ (squares). } 
   \flabel{inertia}
\end{figure}

To interpret our data for the non-classical moment of inertia, we observe
that in an infinite homogeneous system, at the Kosterlitz--Thouless
transition, the superfluid density develops a universal jump\cite{Nelson},
$\Delta \rhos =2 mT_{\KTB}/\pi$, and the superfluid
mass and the moment of inertia are both discontinuous.  In the trap,
the spatial structure smears out these discontinuities \cite{TKTB},
but in local-density approximation, as mentioned, the gas is
critical at the critical radius $\rc$.  Therefore, we expect a normal
phase beyond $\rc$, and a superfluid for $r < \rc$,
with a jump of the superfluid density taking place at this radius and the
superfluid density vanishing for $r>\rc$.  For our parameters,
the superfluid fraction at the critical radius is $\rhos(\rc)/n(\rc)
=2m T /n(\rc)\pi \simeq 0.5$. We can continue 
the superfluid
density $\rhos(r)$ into the inner region by a Thomas--Fermi profile:
\begin{equation}
\rhos(r)= 
\begin{cases}
m \omega^2 r_c^2 \left( 1- r^2/r_c^2 \right)/2g
+ 2 m T/\pi, 
\quad \text{for $r\le r_c$}\\
2m T/\pi \quad \text{for $r\rightarrow  \rc^-$}\\
0 \quad \text{for $r> \rc$}
\end{cases}
\elabel{ansatz}
\end{equation}
(see \fig{profile_0.5}).  The non-classical moment of inertia 
$I_{\nonc}= \int \text{d} {\rvec}\, r^2 [n(r) - \rhos(r)]$, computed
using \eq{ansatz} and the computed density profile $n(r)$,
agrees excellently with our data (see \fig{inertia}).

In the low-temperature phase of a two-dimensional superfluid, the
condensate fraction vanishes in the thermodynamic limit.  In a
homogeneous system, the groundstate has zero momentum, and the fraction
of particles occupying this state can be computed from the long-distance
behavior of the non-diagonal one-body density matrix, $\rho^{(1)}(\rvec,
\rvec';\beta)$ (see  \cite{ceperley}). In an inhomogeneous system,
the groundstate eigenfunction of the one-body density matrix is no
longer completely determined by symmetry.  Still, in the rotationally
symmetric trap the one-body density matrix is block-diagonal with respect
to the Fourier components $l$ of the angle between $\rvec$ and $\rvec'$.
Projection onto the Fourier components yields one-dimensional matrices,
$\rho^{(1)}_l(r,r';\beta)$, which can be  discretized more easily
than the bigger matrix $\rho^{(1)}(\rvec,\rvec';\beta)$.
The condensate fraction, $N_0/N$, corresponds to the largest eigenvalue with $l=0$.
Condensate wavefunctions computed this way for
the  three-dimensional trapped Bose gas, with Legendre polynomials
replacing the Fourier components, closely agree with the solution of
the three-dimensional Gross-Pitaeveskii equation (see \cite{thesis}).
\figg{inertia} shows that the condensate fraction of our quasi
two-dimensional system is rather large, but it decays algebraically with
system size: $N_0/N \sim N^{-\eta(T)/2}$.  The exponent $\eta(T)$
depends on the temperature; we obtain $\eta(0.70\,\TBECZERO)
\approx 0.5$
 and $\eta(0.67\,\TBECZERO) \approx 0.2$.
Precisely at the critical temperature, we expect
$\eta(T_{\KTB})\simeq 1/4$  which implies that the critical
temperature is between $0.67\,\TBECZERO$ and $0.70\,\TBECZERO$,
comparable with our previous estimate, $T_{\KTB} \simeq 0.70\,\TBECZERO$,
based on the occurence of a non-classical moment of inertia. 

Our results are in qualitative agreement with  recent experiments
\cite{Dalibard_2007}. However, they found a lower 
critical temperature $T_{\KTB}^{\text{exp}} = 0.46(3)\,\TBECZERO$, and
an almost Gaussian shape of the density profile above $T_{\KTB}$.
In the specific experiment, the presence of several planes in the
optical lattice renders the
estimation of the critical temperature and the particle numbers difficult.
More generally, we see from \fig{profile_1}, even in the normal state
away from  $T_{\KTB}$, that the density profile deviates from the classical
Gaussian distribution as soon as $n(r)\lambda^2 \gtrsim
1$. Therefore, only a small
part of the density distribution can be used to gauge the temperature
close to $T_{\KTB}$. Fitting a larger part of the distribution 
to a Gaussian typically yields a smaller width than the classical 
thermal distribution and therefore underestimates the true temperature of the
system.
This problem is even
more pronounced when one analyzes column densities rather than
radial profiles.

Notably, in the trap, the universal jump in the superfluid density
of a two-dimensional superfluid at the critical temperature does not
induce a significant discontinuity in the inertial response, as in two-dimensional films
\cite{Reppy}. Nevertheless,
 the universal jump  determines $\rhos(r)$
at the radius $r=\rc$ where the gas is locally critical.
Based on the local-density approximation, 
we proposed a
superfluid density profile,
\eq{ansatz}, which continues the
$\rhos(r)$ from $r=\rc$ into the superfluid inner region. 
It depends 
on a single parameter $r_c$
whose value can be determined directly from the density profile.
The non-classical moment of inertia calculated from the
superfluid density profile, \eq{ansatz},
is in excellent agreement with a direct computation of this quantity.
Further, it is remarkable that in the finite Kosterlitz--Thouless
system, the  condensate fraction, which must vanish for an infinite
system, is still rather large, even close to the transition temperature.
The fact that the groundstate wavefunction of size $\sim \rc$
remains macroscopically occupied
for systems with particle number $N \lesssim 10^6$
implies that the coherence of the atoms
is neither destroyed by interparticle interactions nor
by fluctuations, essential for building
continuous and coherent sources of matter waves
in lower dimensions \cite{Guery_Odelin}.
 
\acknowledgments 
We thank G. Baym, D. Ceperley, I. Cirac, J. Dalibard, and D. Gu\'{e}ry-Odelin for
helpful discussions. M. H. acknowledges support from a CNRS-UIUC exchange
grant.

\end{document}